\documentclass[aps,prb,showpacs,twocolumn,amsmath,amssymb,superscriptaddress]{revtex4} 
\usepackage{graphicx} 
\usepackage{Jonasmacros} 
\usepackage{bm}
\usepackage{mathptmx}
\usepackage{pxfonts}
\usepackage{epstopdf} 
\usepackage{wasysym}

\usepackage[colorlinks,citecolor=blue,linkcolor=magenta]{hyperref}
\usepackage[usenames,dvipsnames,svgnames]{xcolor}

\begin{document}
\date{\today}

\title{Theory of spin inelastic tunneling spectroscopy for superconductor-superconductor and superconductor-metal junctions}

\author{P. Berggren}
\author{J. Fransson}
\email{jonas.fransson@physics.uu.se}
\affiliation{Department of Physics and Astronomy, Uppsala University, Box 516, SE-75120, Uppsala, Sweden}

\begin{abstract}
We address the tunneling conductance and spin inelastic tunneling spectroscopy of localized paramagnetic moments in a superconducting environment, pertaining to recent measurements on Fe-octaethylporphyrin-chloride using superconducting scanning tunneling microscopy. 
We demonstrate that the Cooper pair correlations in the tip and substrate generate a finite uniaxial anisotropy field acting on the local spin moment, and we argue that this field may be a source for the observed changes in the conductance spectrum for decreasing distance between the scanning tunneling tip and the local magnetic moment.
We make a side-by-side comparison between the superconductor-superconductor junction and normal-metal--superconductor junction, and find qualitative agreement between the two setups while quantitative differences become explicit.
 When simulating the effects of electron pumping, we obtain additional peaks in the conductance spectrum that can be attributed to excitations between higher-energy spin states. The transverse anisotropy field couples basis states of the local spin which opens for transitions between spin states that are otherwise forbidden by conservation of angular momentum. Finally, we explore the influences of temperature, which tend to enable in-gap transitions, and an external magnetic field, which enables deeper studies of the spin excitation spectrum. We especially notice the appearance of a low and high excitation peak on each side of the main coherence peak as an imprint of transitions between the Zeeman split ground states.
\end{abstract}

\maketitle
\section{Introduction}
Research into single-spin manipulation remains one of the most active areas in materials science. This is justifiable, as control of single spins would enable information storage with an order-of-magnitude increased density as well as the possible realization of practical quantum computers.\cite{bogani2008,kahle2012} 
Writing and reading information from a localized atomic or molecular spin necessarily involves controlled transitions between different energy states. In most experimental cases under ambient conditions, however, spontaneous interaction with surrounding spin carriers severely limits the mean free lifetime of the local spin excitations below a realistic clock cycle.  

Magnetic atoms or molecules resting on a metal surface, for example, typically deexcite within picoseconds as energy and angular momentum are transferred to itinerant electrons of the substrate. \cite{heinrich2004, balashov2009, khajetoorians2011} As a measure to increase such lifetimes, by limiting the number of ways the local spin can give away energy, an insulating layer can be applied between the metal and the local spin. CuO, BN, and Cu$_2$N have all been used in this manner, to effectively create a gapped substrate, which increases the mean lifetime to hundreds of picoseconds.\cite{kahle2012, lothNP2010, tsukahara2009, loth2012} 

While several novel ways to increase spin excitation lifetimes have been suggested and proven successful, a natural progression from a separating insulating layer is to use a superconducting substrate that exhibits a perfect band gap yet still conducts charge.\cite{miyamachi2013, Christle2014} At low temperatures, a spontaneous deexcitation of the local spin state must then provide enough energy to break up a Cooper pair in order for the main deexcitation mechanism to occur, i.e., quasiparticle-hole pair creation.\cite{lothNP2010} A drawback of a superconducting substrate is the appearance of unwanted Shiba states, within the superconducting gap, generated by exchange interaction between the localized spin moment and the electrons in the superconductor.\cite{yazdani1997, ji2008, franke2011,soda1967,shiba1968} To minimize the effect of these states Heinrich \textit{et al.} successfully utilized a paramagnetic organic  molecule, e.g., M-octaethylporphyrin-chloride (M-OEP-Cl) where M denotes a transition-metal element (Mn, Fe, Co, Ni, Cu), to encage the local magnetic moment such that the direct interaction is kept to a minimum.\cite{Heinrich2013, chen2008} The Shiba states then migrate close to the main coherence peaks and are indiscernible unless the temperature is very low. In addition to providing separation, the ligand cage of the paramagnetic molecule also generates an environment of magnetocrystalline anisotropy for the central magnetic moment that splits up the otherwise degenerate spin states into different energy levels.\cite{tsukahara2009, hirjibehedin2007} This method prolonged the mean lifetime of the first excitation to $\tau\approx10{\rm ns}$, which is enough to clearly observe pumping into higher spin states.\cite{Heinrich2013} The experiments, conducted on a Pb substrate using a Pb covered tip at 1.2 K, shows, in addition, that inelastic scattering between the tunneling electrons and the local spin moment only give signatures in the $dI/dV$ spectra for bias potentials $|eV|=\Delta_{\rm sub}+\Delta_{\rm tip}+\Delta_{mn}$, where $\Delta_{mn}$ is the spin state excitation energy.  
 
\begin{figure}[t]
\begin{center}
\includegraphics[width=0.75\columnwidth]{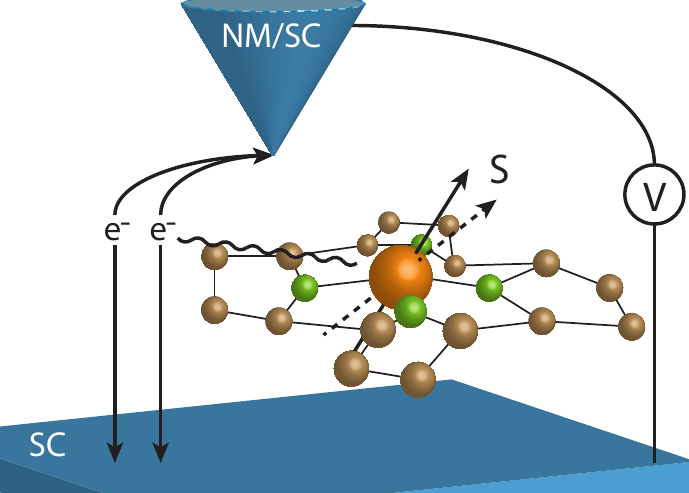}
\caption{(Color online)Schematic illustration of the STM setup. The tip is modeled both as a normal metal and a superconductor, while only a superconducting substrate is considered. As indicated by the arrows, electrons may tunnel directly between tip and substrate or through intermediate interaction with the local spin $S$, of the paramagnetic molecule, under bias voltage $V$.}
\label{fig0}
\end{center}
\end{figure} 

The theoretical model derived in this paper emulates single-electron tunneling in a scanning tunneling microscope (STM) setup where the tip is made up of a normal metal (NM) or a superconductor (SC). As a substrate, on which a paramagnetic organic molecule lies, only a superconductor is considered. The magnetic center of the molecule provides a local spin moment, within an anisotropic environment, elevated enough to prevent significant direct magnetic interaction with the close-by superconductors. An applied bias voltage, which controls the relative Fermi levels of the tip and substrate, will induce a tunneling current of electrons that either pass the local spin moment unnoticed or interact with exchange of energy and angular momentum. See Fig. \ref{fig0} for a sketched illustration of the setup.      

In excellent agreement with experiment, our transparent (differential) conductance expression yields signatures of inelastic spin transitions only outside of the tip and substrate superconducting gap at low temperatures. We also reproduce the observed effects of pumping to reveal interactions with higher spin states. Beyond the reproduction of experimental results, the conductance spectra are thoroughly investigated with respect to varying anisotropies and external magnetic fields.  

In particular, we investigate the effect of the Cooper pair correlations in the tip and substrate on the spectrum of the local magnetic moments. We show that these give rise to an additional contribution to the uniaxial anisotropy, something which can partially explain the observed spectral changes upon bringing the STM tip closer to the paramagnetic sample. \cite{Heinrich2013,heinrich2014}

While some of the results regarding the conductance spectrum for the SC-SC junction have been published elsewhere,\cite{berggren2014} the present paper also includes the above-mentioned investigation of the Cooper-pair-induced uniaxial anisotropy field, the case of a localized magnetic moment embedded in a normal-metal--superconductor junction, and a systematic study of the influence of temperature and external magnetic fields. For completeness and to enable a comparison between the different scenarios, we also include details of the SC-SC junction.

The paper is organized as follows. In Sec. \ref{sec-model} we define the microscopical model for our setup, in Sec. \ref{sec-modifications} we elucidate the impact of the tip and substrate electrodes and the tunneling current on the spin excitation spectrum, in Sec. \ref{sec-current} we derive the expressions for the tunneling current including the expression pertaining to the inelastic electron tunneling spectroscopy (IETS) measurements, in Sec. \ref{sec-analysis} we present and analyze the main results for spin $S=1$ and $S=5/2$ systems, and we conclude the paper in Sec. \ref{sec-summary}.

\section{Theoretical description of the model}
\label{sec-model}
The electronic composition and interplay within the STM device is governed by the total Hamiltonian 
\begin{align}
\Hamil=&
	\Hamil_{\rm tip}+\Hamil_{\rm sub}+\Hamil_{{\rm T}}+\Hamil_{\rm S}
		+\Hamil_{Kt}+\Hamil_{Ks}
	,
\label{eq-Hamil}
\end{align} 
where $\Hamil_{\rm tip}$ and $\Hamil_{\rm sub}$ give the electronic structure of the tip and the substrate, respectively. In this study, we consider two different types of tip states: (i) normal metal and (ii) superconducting. These are modeled using
\begin{subequations}
\label{eq-tip}
\begin{align}
\Hamil_{tip}^{(NM)}=&
	\sum_{\bfp\sigma}\varepsilon_{\bfp\sigma}\cdagger{\bfp}\cc{\bfp}
	,
\label{eq-NMtip}
\\
\Hamil_{\rm tip}^{(SC)}=&
	\sum_{\bfp\sigma}\varepsilon_{\bfp\sigma}c^\dagger_{\bfp\sigma}c_{\bfp\sigma}+\sum_{\bfp\sigma}\Delta_{\rm tip}c^\dagger_{\bfp\uparrow}c^\dagger_{-\bfp,\downarrow}+\mbox{H.c.}
	,
\label{eq-SCtip}
\end{align}
\end{subequations}
respectively. Here, $c^\dagger_{\bfp\sigma}$ $(c_{\bfp\sigma})$ creates (destroys) an electron/quasiparticle in the tip with momentum $\bfp$ and spin $\sigma=\up,\down$. The substrate quasiparticles are modeled by
\begin{align}
\Hamil_{\rm sub}=&
	\sum_{\bfk\sigma}\varepsilon_{\bfk\sigma}c^\dagger_{\bfk\sigma}c_{\bfk\sigma}+\sum_{\bfk\sigma}\Delta_{\rm sub}c^\dagger_{\bfk\uparrow}c^\dagger_{-\bfk,\downarrow}+\mbox{H.c.}
,
\label{eq-sub}
\end{align}
where $c^\dagger_{\bfk\sigma}$ $(c_{\bfk\sigma})$ and $\bfk$ denote electron operators and momentum, respectively.
The three parts, tip, substrate, and sample, are connected via tunneling through
\begin{align}
\Hamil_{\rm T}=&
	\sum_{\substack{\bfp\bfk\\\sigma\sigma'}}
		\cdagger{\bfp}\left[T_0\delta_{\sigma\sigma'}+T_1\bfsigma_{\sigma\sigma'}\cdot\bfS\right]\cc{\bfk\sigma'}+\mbox{H.c.}.
\end{align}
One direct tunneling path, with a rate $T_0$ ($\delta_{\sigma\sigma'}$ is the Kronecker delta), is spin preserving, whereas a second tunneling path, with rate $T_1$, accounts for the interaction between the electron spin and localized spin. Here, $\bfsigma_{\sigma\sigma'}$ is the Pauli-matrix vector. The ratio $T_1/T_0$ may be of the order of unity since this rate is determined both by the tunneling overlap as well as the Coulomb assisted tunneling rate; see, e.g., Ref. \onlinecite{patton2007}.

In addition to the tunneling contribution $\Hamil_T$, we consider the effects of Kondo like coupling between the local spin moment and the electrons in the tip and substrate, respectively. This is introduced through the contributions
\begin{align}
\Hamil_{Kt}+\Hamil_{Ks}=&
	\sum_{\sigma\sigma'}
	\biggl(
		T_t\sum_{\bfp\bfp'}\cdagger{\bfp}\bfsigma_{\sigma\sigma'}\cc{\bfp'\sigma'}
		+
		T_s\sum_{\bfk\bfk'}\cdagger{\bfk}\bfsigma_{\sigma\sigma'}\cc{\bfk'\sigma'}
	\biggr)
	\cdot\bfS,
	\label{eq:Kondolike}
\end{align}
where $T_{t(s)}$ is the energy for the coupling between the local spin moment and the electrons in the tip (substrate).

The local magnetic moment, embedded in the anisotropic environment of the organic molecule, derives its $2S+1$-fold spectrum of spin eigenenergies and states $\{E_\alpha,|\alpha\rangle\}$ from the Hamiltonian
\begin{equation}
\Hamil_{\rm S}=
	-g\mu_B\bfB\cdot\bfS
	+DS_z^2
	+\frac{E}{2}(S_+^2+S_-^2).
\label{eq:Hs}
\end{equation}
Here, $g$ is the gyromagnetic ratio, $\mu_B$ is the Bohr magneton, and $\bfB$ is an external magnetic field.\cite{Gatteschi2006} For arbitrary integer total spin moment $S$ and a finite uniaxial anisotropy field $D$, the basis states $|S_z,m_z\rangle$, $m_z=-S_z,-S_z+1,\ldots,S_z$, remain eigenstates with twofold degenerate excitations. The transverse anisotropy field $E$ will split up these excitations as well as cause the eigenstates to form linear combinations of the basis states. A half-integer spin moment behaves much in the same way under finite anisotropies, with the exception that $E$ no longer splits up the two fold degenerate excitations but rather shifts the energy levels somewhat.  

Direct local interactions with the substrate, responsible for Shiba states, are described by $\Hamil_{Ks}$. For a large local spin moment $(S\gg1)$ that couples weakly $(T_s\rightarrow0)$ to the surface electrons (ensured by the separating ligand cage), in-gap resonances appear at energies $\omega_0=\pm\Delta_{\rm sub}[1-(\pi NT_sS/2)^2]/[1+(\pi NT_sS/2)^2]$,\cite{shiba1968} where $N$ is the substrate density of states (DOS). The energies of the Shiba states, hence, approach the edges, or coherence peaks, of the superconducting gap whenever $(\pi NJS/2)^2\ll1$.

In a quantum mechanical treatment, the energies of the Shiba states have to be considered using perturbation theory. To the lowest-order approximation, the energies of the Shiba states are given by $\Delta_{\rm sub}[1-\alpha(NJS/2)^2]$, where $\alpha=9$ ($\alpha=1$) for antiferromagnetic (ferromagnetic) coupling $J$. Despite the different form of the excitation energies obtained in this perturbational approach, the general statement that the energies approach the edges of the superconducting gap in the limit $(\pi NJS/2)^2\ll1$ remains valid.

We hence conclude that for $NJS<2k_BT/3\pi$, the Shiba are hidden within the thermally broadened coherence peaks. Although these states may influence the lifetime of the spin excitations, we omit this contribution in the following discussion in order to focus on the signatures observed in experiment.
 
\section{Modifications to the spin spectrum induced by tip, substrate, and current}
\label{sec-modifications}
Here we discuss a possible reason and explanation for the strong changes in the magnetic anisotropy, acting on the local magnetic moment as a function of the tunneling current, which was observed in Refs. \onlinecite{Heinrich2013,heinrich2014}.

From the microscopical model introduced in Eq. (\ref{eq-Hamil}), we can derive an effective spin Hamiltonian which is suitable for studies of possible sources of broadening and level shift in the spin excitation spectrum. Employing the methods used in Refs. \onlinecite{zhu2004,fransson2008,bhattacharjee2012,fransson2014}, we arrive at the effective model for the local spin,
\begin{align}
\Hamil_S^{eff}=&
	\Hamil_S
	+{\cal D}\bfS\cdot\bfS
	+2{\cal F}S_z^2,
\label{eq-effHs}
\end{align}
where both fields ${\cal D}$ and ${\cal F}$ comprise three contributions arising from the local spin interactions with electrons in the tip (${\cal D}_t/{\cal F}_t$), substrate (${\cal D}_s/{\cal F}_s$), and tunneling current (${\cal D}_c/{\cal F}_c$). We notice that the second contribution to the spin spectrum, ${\cal D}\bfS\cdot\bfS={\cal D}S^2$, does not provide any essential changes to the relative level spacing in the spectrum, except for an overall shift of the energy. As we shall see, however, this field contributes a source of dissipation whenever the voltage bias is greater than $2|\Delta|$, which generates a current-induced broadening of the spin excitations. Second, we notice that the presence of the superconducting electrodes and tunneling current generates a contribution to the axial anisotropy in the last term of Eq. (\ref{eq-effHs}), which was also observed in experiments. \cite{Heinrich2013,heinrich2014}

The contribution from the current mediated interactions is given by
\begin{subequations}
\begin{align}
{\cal D}_c=&
	-{\cal F}_c
\nonumber\\&
	-2T_1^2
	\sum_{\bfp\bfk}
	\int
		\frac{G_\bfp^<(\omega)G^>_\bfk(\omega')-G^>_\bfp(\omega)G^<_\bfk(\omega')}
			{\omega-\omega'+eV-i\delta}
	\frac{d\omega}{2\pi}\frac{d\omega'}{2\pi}
	,
\label{eq-calD}
\\
{\cal F}_c=&
	-2T_1^2
	\sum_{\bfp\bfk}
	\int
		\frac{F^{+<}_\bfp(\omega)F^>_\bfk(\omega')-F^{+>}_\bfp(\omega)F^<_\bfk(\omega')}
			{\omega-\omega'-eV-i\delta}
	\frac{d\omega}{2\pi}\frac{d\omega'}{2\pi}
	e^{-i2eVt}
	.
\label{eq-calF}
\end{align}
\end{subequations}
The tip (substrate) contribution is obtained analogously in the limit $V\rightarrow0$ by replacing $\bfk$ ($\bfp$) with $\bfp'$ ($\bfk'$). Here, we have introduced the lesser anomalous surface Green functions GFs $F^<_\bfk(\omega)=i\av{\cc{\bfk\up}\cc{-\bfk\down}}(\omega)$ and $F^{+<}_\bfk(\omega)=i\av{\csdagger{-\bfk\down}\csdagger{\bfk\up}}(\omega)$, the corresponding greater GFs $F^>_\bfk(\omega)=(-i)\av{\cs{-\bfk\down}\cc{\bfk\up}}(\omega)$ and $F^{+>}_\bfk(\omega)=(-i)\av{\csdagger{\bfk\up}\csdagger{-\bfk\down}}(\omega)$, as well as the corresponding lesser and greater anomalous GFs for the tip in the case of a superconducting tip. Otherwise, these propagators are absent for the tip, which implies that ${\cal F}_{t/c}\equiv0$.

Before moving ahead, we notice, however, that the field ${\cal F}_c$, which is generated by the Cooper pair correlations in the tunneling current, acquires a harmonic temporal variation with the ac Josephson current and, hence, vanishes on average. The spin excitation spectrum is, therefore, expected to fluctuate around its static spectrum with a frequency given by the Josephson current. For the IETS measurements in this paper, we will therefore neglect the influence of this anisotropy field as the tunneling spectroscopy is obtained in the long-time limit.

\begin{figure}[t]
\begin{center}
\includegraphics[width=0.99\columnwidth]{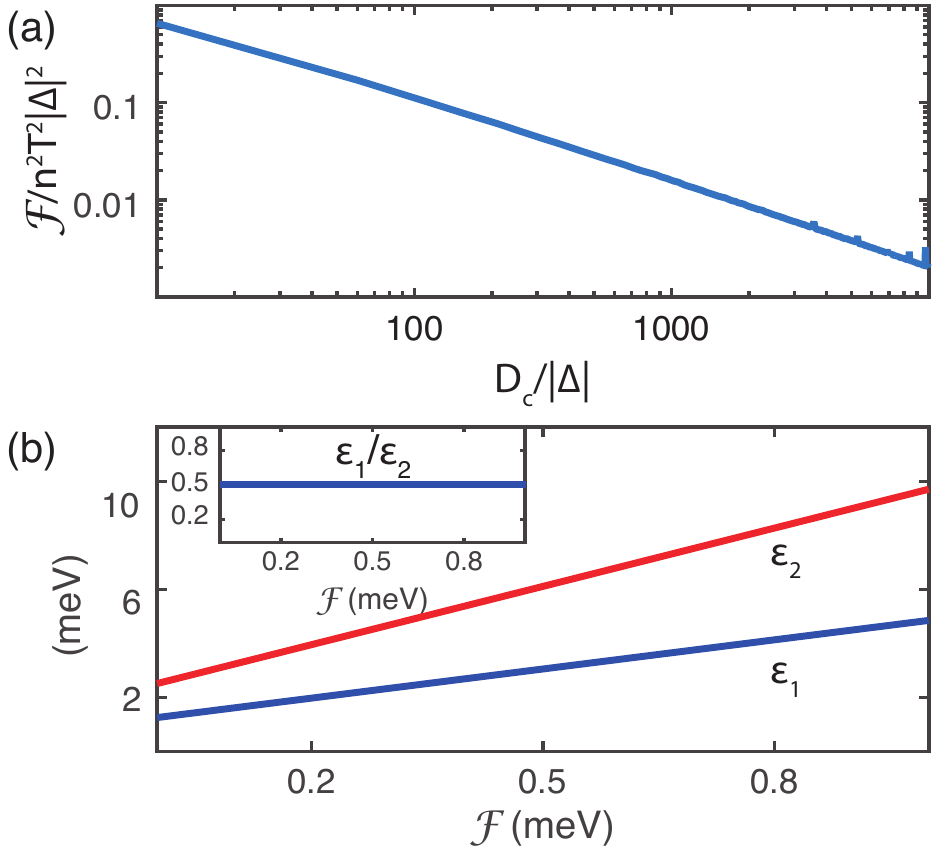}
\end{center}
\caption{(Color online) (a) Typical dependence of the anisotropy energy ${\cal F}_{t/s}$ as a function of the high-energy cutoff $D_c$. (b) Transition energies $\dote{1}=E_{3/2}-E_{1/2}$ and $\dote{2}=E_{5/2}-E_{3/2}$ as a function of ${\cal F}$ for a $S=5/2$ local moment with uniaxial anisotropy $D=0.7$ meV and $E=0$. Inset: The ratio between the two transition energies.}
\label{fig-DF}
\end{figure}

The tip and the substrate are assumed to be in local equilibrium which allows us to employ the fluctuation-dissipation theorem. For a general lesser/greater GF ${\cal A}^{</>}$, we can then make use of the relation ${\cal A}^{</>}(\omega)=(\pm i)f(\pm\omega)[-2\im{\cal A}^r(\omega)]$, where ${\cal A}^r$ is the corresponding retarded propagator. Although the effect of ${\cal F}_c$ is negligible in the finite voltage bias regime, we write the full expression here for completeness, and also since ${\cal F}_{t/s}$ are easily obtained from these expressions in the limit $V\rightarrow0$. We have
\begin{align}
{\cal F}_c=&
	\frac{2T_1^2}{\pi}
	e^{-i2eVt}
	\sum_{\bfp\bfk}
	\int
	\Biggl(
		f(\omega)
		\im\Bigl[
			F^{+r}_\bfp(\omega)F^r_\bfk(\omega-eV)
		\Bigr]
\nonumber\\&
	-
		\Bigl(
			f(\omega)-f(\omega-eV)
		\Bigr)
		F^{+r}_\bfp(\omega)
		\im F^r_\bfk(\omega-eV)
	\Biggr)
	d\omega
	.
\end{align}
Since the effect of ${\cal F}$ in the finite voltage bias regimes is discarded, we write ${\cal D}$ according to ${\cal D}={\cal D}_0+{\cal D}_V$, where the subscript 0 ($V$) refers to zero (finite) voltage bias. The contribution from the tunneling current generates
\begin{subequations}
\label{eq-calDD}
\begin{align}
{\cal D}_{c0}=&
	\frac{2T_1^2}{\pi}
	\im
	\sum_{\bfp\bfk}
	\int
		f(\omega)G^r_\bfp(\omega)G^r_\bfk(\omega)
	d\omega
	-
	{\cal F}_0
	,
\label{eq-calD0}
\\
{\cal D}_{cV}=&
	\frac{2T_1^2}{\pi}
	\sum_{\bfp\bfk}
	\int
	\Biggl(
		f(\omega)
		\im\Bigl[
			G^r_\bfp(\omega)G^r_\bfk(\omega-eV)
		\Bigr]
\nonumber\\&
		-
		\Bigl(
			f(\omega)-f(\omega-eV)
		\Bigr)
		G^r_\bfp(\omega)\im G^r_\bfk(\omega-eV)
	\Biggr)
	d\omega,
\label{eq-calDV}
\end{align}
\end{subequations}
where ${\cal F}_0$ is the zero voltage bias result. Here, we notice a few interesting issues with the current-induced anisotropy field ${\cal D}$. In equilibrium, $V=0$, the field ${\cal D}_0$ is purely real which implies that it merely provides an overall energy shift of the spin spectrum. Moreover, in the nonsuperconducting case, the expression for ${\cal D}_0$ reproduces the electronically mediated exchange interactions between spin moments at different points in space and/or time; e.g., see Refs. \onlinecite{antropov1995,antropov1997,katsnelson2004,bhattacharjee2012,fransson2014}. Here, since we address a system with a single spin, this interaction can be understood as a self-interaction.

We proceed with the calculations of the fields ${\cal D}$ and ${\cal F}$ by introducing the retarded GF for the surface electrons in the substrate which, in Nambu space, can be written as
\begin{align}
\bfG_\bfk^r(\omega)\equiv&
	\begin{pmatrix}
		G^r_\bfk & F^r_\bfk \\ F^{+r}_\bfk & G^{+r}_\bfk
	\end{pmatrix}
	(\omega)
\nonumber\\=&
	\frac{1}{(\omega+i\delta)^2-E_\bfk^2}
	\begin{pmatrix}
		\omega+i\delta+\dote{\bfk} & \Delta \\ \Delta^* & \omega+i\delta-\dote{\bfk}
	\end{pmatrix}
	,
\end{align}
and analogously for the electrons in the tip. Summing over the momentum, assuming energy independent density of electron states $n_{\rm sub}$, we then have
\begin{subequations}
\begin{align}
G^r_{\rm sub}(\omega)=&
	-n_{\rm sub}
	\frac{\omega}{\sqrt{\omega^2-|\Delta_{\rm sub}|^2}}
	\theta(|\omega|-|\Delta_{\rm sub}|)
	\Lambda_{\rm sub}(\omega),
\\
F^r_{\rm sub}(\omega)=&
	-n_{\rm sub}\frac{\Delta_{\rm sub}}{\sqrt{\omega^2-|\Delta_{\rm sub}|^2}}
	\Lambda_{\rm sub}(\omega)
\end{align}
\end{subequations}
and analogously for the tip GF. Here, we have introduced the notation $\Lambda_{\rm sub}(\omega)=\ln|1-2\omega_{\rm sub}/(\omega_{\rm sub}+D_c)|+i\pi{\rm sgn}(\omega)/2$ where $\omega_{\rm sub}=\sqrt{\omega^2-|\Delta_{\rm sub}|^2}$ and $D_c$ is a high-energy cutoff, or half the width of the metallic band.

In equilibrium, the Cooper pair correlations then give rise to the anisotropy from, e.g., the tip,
\begin{align}
{\cal F}_s=&
	n_{\rm sub}^2T_1^2
	\int
		f(\omega)
		\frac{|\Delta_{\rm sub}|^2{\rm sgn}(\omega)}
		{\omega^2-|\Delta_{\rm sub}|^2}
		\ln\biggl|1-2\frac{\omega_{\rm sub}}{\omega_{\rm sub}+D_c}\biggr|
	d\omega.
\end{align}
Although the integrand in this expression is logarithmically small, it is non-negligible due to the integration over occupied states, which essentially spans the energy range $[-D_c,0]$. We can, moreover, see that this energy is finite, although decreasing with the high-energy cutoff, which is illustrated in Fig. \ref{fig-DF}(a). Hence, for a given bandwidth $D_c$, the anisotropy constant $\propto (T_sn_{\rm sub}|\Delta_{\rm sub}|)^2$ is essentially determined by the coupling strength $T_s$, which varies exponentially with the distance between the local spin moment and the substrate, and analogously for the local moment and the tip.

Here, we make contact with the experiments performed in Ref. \onlinecite{Heinrich2013}, which involves a $S=5/2$ local moment, where the STM tip to molecular sample distance was varied in the tunnel junction. As an effect, the transition energies drifted towards higher energies with decreasing distance resulting from an increased uniaxial anisotropy field acting on the local spin moment. By solving the effective spin Hamiltonian given by Eq. (\ref{eq-effHs}) as a function of ${\cal F}$ for, e.g., a $S=5/2$ local moment, we obtain the spectrum $E_{n/2}=(n/2)^2(D+{\cal F})$, which gives the transition energies $\dote{1}=E_{3/2}-E_{1/2}=2(D+2F)$ and $\dote{2}=E_{5/2}-E_{3/2}=4(D+2F)$, as shown in Fig. \ref{fig-DF}(b), as well as a constant ratio $\dote{1}/\dote{2}=1/2$; see inset of Fig. \ref{fig-DF}(b). From this, we conclude that although the proximity mechanism of the superconducting tip and substrate may not be the sole cause of an increased anisotropy with decreasing the tip-sample distance, the Cooper pair correlations within, e.g., the superconducting tip create a finite source for the anisotropy. As the effect from the Cooper pair correlations is essentially constant, which is crucial in the context, the increased anisotropy field with decreased tip-sample distance is provided by the exponential distance dependence in the coupling parameter $T_{t(s)}$.

Under nonequilibrium conditions, the modification of the spin excitation spectrum due to the tunneling current is not well understood at the moment since the temporally fluctuating field induced by the Josephson current vanishes on average, but also since the field induced by the Josephson current is expected to decay with increasing voltage bias. Moreover, the field ${\cal D}_{cV}$ only acts on $S^2$ and, hence, merely provides a rigid shift of the spectrum. We, therefore, proceed with our discussion about the tunneling current and conductance.

\section{The tunneling current}
\label{sec-current}
The tunneling current $I(t)$ is derived from the relation
\begin{align}
I(t)=&
		-e\partial_t\sum_{\bfp\sigma}\av{\cdagger{\bfp}\cc{\bfp}}
	=
		i2e\im\sum_{\bfp\sigma}\av{\cdagger{\bfp}\com{\cc{\bfp}}{\Hamil_T}},
\label{eq:currentnr1}
\end{align}
with the electronic charge $e$ and $\hbar=1$, which leads to
\begin{widetext}
\begin{align}
I(t,V)=&
	2e\mbox{Re}\sum_{\substack{\mathbf{p}\mathbf{k}\\\sigma\sigma'}}\sum_{\substack{\mathbf{p'}\mathbf{k'}\\\sigma''\sigma'''}}\int^t_{-\infty}\Big\langle\Big[c^\dagger_{\mathbf{p}\sigma}(t)\hat{T}_{\sigma\sigma'}(t)c_{\mathbf{k}\sigma'}(t),c^\dagger_{\mathbf{p'}\sigma''}(t')\hat{T}_{\sigma''\sigma'''}(t')c_{\mathbf{k'}\sigma'''}(t')\Big]e^{ieV(t+t')}\\&
		+\Big[c^\dagger_{\mathbf{p}\sigma}(t)\hat{T}_{\sigma\sigma'}(t)c_{\mathbf{k}\sigma'}(t),c^\dagger_{\mathbf{k'}\sigma'''}(t')\hat{T}^\dagger_{\sigma''\sigma'''}(t')c_{\mathbf{p'}\sigma''}(t')\Big]e^{ieV(t-t')}\Big\rangle dt',
\end{align}
\end{widetext}
where $\hat{T}_{\sigma\sigma'}(t)=T_0\delta_{\sigma\sigma'}+T_1\boldsymbol{\sigma}_{\sigma\sigma'}\cdot\mathbf{S}_n(t)$, whereas $eV$ represents the applied bias voltage by letting $\varepsilon_{\bfp\sigma}\rightarrow\varepsilon'_{\bfp\sigma}+eV/2$ and $\varepsilon_{\bfk\sigma}\rightarrow\varepsilon'_{\bfk\sigma}-eV/2$. The two terms inside the expectation value account for two different tunneling processes, of which the first gives the Josephson tunneling contribution and will be omitted throughout the rest of the paper since we are interested in the regime far from equilibrium where the Josephson correlations are negligible. The second term describes single-electron tunneling and may, analogously to previous studies on NM leads, be divided into three different parts proportional to $I_0\propto T_0^2$, $I_1\propto T_0T_1$, and $I_2\propto T_1^2$.\cite{fransson2009, fransson2010, lorente2009, rossier2009, persson2009} 

$I_0$ is the direct tip to substrate tunneling current,
\begin{align}
I_0(t,V)=&
	2eT_0^2\mbox{Re}\sum_{\bfp\bfk\sigma}\int^t_{-\infty}\bigg[G^<_{\mathbf{p},\sigma}(t',t)G^>_{\bfk,\sigma}(t,t')
\nonumber\\&
	-G^>_{\bfp,\sigma}(t',t)G^<_{\bfk,\sigma}(t,t')\bigg]e^{ieV(t-t')}dt',
\label{eq:spincurrent0}
\end{align}
and contains electron Green functions of the kind $G^<_{\mathbf{p},\sigma}(t',t)=i\langle \cdagger{\bfp}(t)\cc{\bfp}(t')\rangle_{\Hamil_{\rm tip}}$, where $\Hamil_{\rm tip}$ determines the electron environment in the tip. Contributions from $I_0$ are kept as a background current in our calculated results. $I_1$, on the other hand, does couple to the local spin moment, which can be seen from
\begin{align}
I_1(t,V)=&
	2eT_0T_1\mbox{Re}\sum_{\substack{\bfp\bfk\\\sigma\sigma'}}
		\int^t_{-\infty}\av{\bfsigma_{\sigma\sigma'}\cdot\bfS(t)+\bfsigma_{\bar{\sigma}\bar{\sigma}'}\cdot\bfS(t')}
			e^{ieV(t-t')}
\nonumber\\&\times
	\bigg[G^<_{\mathbf{p},\sigma}(t',t)G^>_{\bfk,\sigma'}(t,t')-G^>_{\bfp,\sigma}(t',t)G^<_{\bfk,\sigma'}(t,t')\bigg]
	dt'
	.
\label{eq:spincurrent1}
\end{align}
This contribution is discarded, however, since it vanishes in the absence of spin-polarized currents.\cite{fransson2009, fransson2010}. $I_2$, finally, contains the spin-spin correlation function of the local magnetic moment coupled to the tunneling current. In general, 
\begin{align}
I_2(t,V)=&
	2eT_1^2\mbox{Re}
	\sum_{\substack{\bfp\bfk\\\sigma\sigma'}}\int^t_{-\infty}\bfsigma_{\sigma\sigma'}\cdot\bigg[G^<_{\mathbf{p},\sigma}(t',t)G^>_{\bfk,\sigma'}(t,t')\bfchi^>(t,t')
\nonumber\\&
	-G^>_{\bfp,\sigma}(t',t)G^<_{\bfk,\sigma'}(t,t')\bfchi^<(t,t')\bigg]\cdot\bfsigma_{\sigma'\sigma}e^{ieV(t-t')}dt',
\label{eq:spincurrent2}
\end{align}
where $\bfchi^>(t,t')=\av{\bfS(t)\bfS(t')}$ and $\bfchi^<(t,t')=\av{\bfS(t')\bfS(t)}$.

The electron-spin projection onto the spin-correlation functions of the local magnetic moment amounts to
\begin{equation}
\bfsigma_{\sigma\sigma'}\cdot\left\langle\bfS(t)\bfS(t')\right\rangle\cdot\bfsigma_{\sigma'\sigma}=\sum_{\alpha\beta}\left(2\chi^z_{\alpha\beta}+\chi^{-+}_{\alpha\beta}+\chi^{+-}_{\alpha\beta}\right)e^{i(E_\alpha-E_\beta)(t-t')},
\label{eq:spincorrel1}
\end{equation}
where 
\begin{align}
\chi^{z,-+,+-}_{\alpha\beta}=&\left\langle\alpha|S^{z,-,+}|\beta\right\rangle\left\langle\beta|S^{z,+,-}|\alpha\right\rangle
	P(E_\alpha)\left[1-P({E_\beta})\right],
\label{eq:spincorrel2}
\end{align}
and the labels $\alpha$, $\beta$ refer to the eigensystem $\{E_\alpha,\ket{\alpha}\}$ for the spin Hamiltonian in Eq. (\ref{eq:Hs}). In the following, we employ the decoupling $\av{(d^\dagger_\alpha d_\beta)(t)(d^\dagger_\beta d_\alpha)(t')}={\cal G}_\beta^>(t,t'){\cal G}^<_\alpha(t',t)$, where ${\cal G}^<_\alpha(t,t')=i\av{d^\dagger_\alpha(t')d_\alpha(t)}$ and ${\cal G}_\alpha^>(t,t')=(-i)\av{d_\alpha(t)d^\dagger_\alpha(t')}$ are the lesser/greater Green functions for the spin excitations. In the atomic limit, we have ${\cal G}^{</>}_\alpha(t,t')=(\pm i)f(\pm E_\alpha)e^{-iE_\alpha(t-t')}$, where $f(\omega)$ is the Fermi function.

As we consider stationary conditions in the setup, we Fourier transform our theory to the energy domain, using, e.g., $g(\omega)=\int g(t-t')e^{i\omega(t-t')}dt'$.

\subsection{Normal metal to superconductor junction}
For a normal-metal (NM) to superconductor (SC) junction, the conditions differ for tip and substrate electrons/quasiparticles and, consequently, so does their respective Green functions. At the outset, considering the tip to be of a NM, the electronic structure is provided by Eq. (\ref{eq-NMtip}). The lesser and greater Green functions of the tip are then simply
\begin{subequations}
\begin{align}
G^{</>}_{\bfp\sigma}(\omega)=&
	(\pm i)f(\pm\omega)\delta(\varepsilon_{\bfp}-\omega),
\label{eq:metalgreen}
\end{align}
\end{subequations}
where $f(\omega)$ is the Fermi-Dirac distribution function. Within the SC substrate, the quasiparticle structure given by Eq. (\ref{eq-sub}) leads to lesser/greater Green functions,
\begin{subequations}
\label{eq-Gsc}
\begin{align}
G^{</>}_{\bfk\sigma}(\omega)=&
	(\pm i)\left\{|u_\bfk|^2f(\pm\omega)\delta(E_{\bfk}+\omega)+|v_\bfk|^2f(\mp\omega)\delta(E_{\bfk}-\omega)\right\},
\end{align}
\end{subequations}
where $u_\bfk=\sqrt{1/2(1+\varepsilon_\bfk/E_\bfk)}$ and $v_\bfk=\sqrt{1/2(1-\varepsilon_\bfk/E_\bfk)}$ are the coherence factors, and the quasiparticle energy $E_\bfk=\sqrt{{\varepsilon}_\bfk^2+|\Delta_{sub}|^2}$.

Following the approach in, e.g., Ref. \onlinecite{mahan1990}, we let $\sum_\bfp\rightarrow\int d\varepsilon \hspace{1mm}n_{\rm tip}$ and $\sum_\bfk\rightarrow\int^\infty_{\Delta_\bfk}dE_\bfk\hspace{1mm}n_{\rm sub} E_\bfk/\sqrt{E^2_\bfk-|\Delta_\bfk|^2}$ in Eqs. (\ref{eq:spincurrent0}) and (\ref{eq:spincurrent2}), where $n_{\rm tip}$ and $n_{\rm sub}$ are the energy-independent tip and substrate electron/quasiparticle density coefficients, respectively, to obtain
\begin{align}
I_0(V)=&4\pi eT_0^2n_{\mathbf{p}}n_{\mathbf{k}}\int^{\infty}_{\Delta_{\mathbf{k}}}
	dEE\frac{f(E-eV)-f(E+eV)}{\sqrt{E^2-|\Delta_{\mathbf{k}}|^2}}
\label{eq:backgcurrent}
\end{align}
for direct tunneling and
\begin{align}
I_2(V)=&
	4\pi eT_1^2n_{\rm tip}n_{\rm sub}\sum_{\alpha\beta}\int^{\infty}_{\Delta_{\mathbf{k}}}
		dEE
		\frac{2\chi^z_{\alpha\beta}+\chi^{-+}_{\alpha\beta}+\chi^{+-}_{\alpha\beta}}{\sqrt{E^2-|\Delta_{\mathbf{k}}|^2}}
\nonumber\\&
	\times\bigg\{f\big(E-E_\alpha+E_\beta-eV\big)-f\big(E-E_\alpha+E_\beta+eV\big)
\nonumber\\&
	+\bigg[f\big(E+E_\alpha-E_\beta-eV\big)-f\big(E+E_\alpha-E_\beta+eV\big)
\nonumber\\&
	+f\big(E-E_\alpha+E_\alpha+eV\big)-f\big(E-E_\alpha+E_\beta-eV\big)\bigg]f\big(E\big)\bigg\}
\label{eq:spinexcurrent}
\end{align}
for spin-exchange tunneling. The remaining energy integrals of the two expressions are solved numerically.  

Equation (\ref{eq:backgcurrent}) provides a qualitative background shape for the current, as a function of bias voltage, that modifies under interaction with the local spin moment when (\ref{eq:spinexcurrent}) is added. Considering positive bias voltages, $f(E+eV)\approx0$ at low temperatures, while $f(E-eV)$ suddenly jumps from 0 to 1 when $E$ and $eV$ start to match. The tunneling current is, consequently, close to 0 while $0\leq eV\leq\Delta_\bfk$; it quickly rises to a value determined by the quotient factor once $eV\approx\Delta_\bfk$ and transitions into a linear increase for $eV>\Delta_\bfk$. In the conductance $dI/dV(eV)$ spectra, the behavior is reflected in a sharp peak structure at $eV=\Delta_\bfk$, preceded by $dI/dV\approx0\hspace{1mm}{\rm A/V}$, and followed by a constant value. 

When the bias voltage is positive, the explicit expression for $I_2(V)$ suggests that a contributing tunneling channel opens up once $eV\geq\Delta_\bfk+E_\beta-E_\alpha$ for transitions allowed by $\chi^z_{\alpha\beta}$, $\chi^{-+}_{\alpha\beta}$, $\chi^{+-}_{\alpha\beta}$, i.e., that conserves angular momentum. In the dI/dV spectrum, this means that the initial peak at current onset is accompanied by smaller peaks for higher voltages corresponding to the spin excitation energies of the possible transitions. Apart from losing energy to the local spin by exciting it from the ground state, a tunneling electron may also gain energy from deexcitation of a thermally populated higher state. Such occurrences cause dI/dV peaks at lower voltages than the main peak. For negative voltages, the dI/dV spectrum is a mirror image with respect to $eV=0$.

\subsection{Superconductor to superconductor junction}
Changing the STM tip from a NM to a SC, that is, using the quasiparticle structure given by Eq. (\ref{eq-SCtip}), we replace the tip GF by the ones given in Eq. (\ref{eq-Gsc}) (replacing $\bfk\rightarrow\bfp$).
Hence, using Eqs. (\ref{eq:spincurrent0}) and (\ref{eq:spincurrent2}), the direct and spin-exchange currents now become
\begin{widetext}
\begin{align}
I_0(V)=&4\pi eT_0^2n_{\rm tip}n_{\rm sub}\int^{\infty}_{\Delta}dE\frac{E}{\sqrt{E^2-|\Delta|^2}}
	\Bigg\{\Big[\theta(E+eV-\Delta)-\theta(-E-eV-\Delta)\Big]\Big[f(E)-f(E+eV)\Big]\frac{E+eV}{\sqrt{(E+eV)^2-|\Delta|^2}}
\nonumber\\&
	+\Big[\theta(-E+eV-\Delta\big)-\theta(E-eV-\Delta)\Big]
	\Big[f(E)-f(E-eV)\Big]\frac{E-eV}{\sqrt{(E-eV)^2-|\Delta|^2}}\Bigg\}
\label{eq:SCSCspinexcurrent0}
\end{align}
for direct tunneling, and
\begin{align}
I_2(V)=&4\pi eT_1^2n_{\rm tip}n_{\rm sub}\sum_{\alpha\beta}\sum_{n=1}^4\int^{\infty}_{\Delta}dE\frac{(-1)^nE}{\sqrt{E^2-|\Delta|^2}}
			\big(2\chi^z_{\alpha\beta}+\chi^{-+}_{\alpha\beta}+\chi^{+-}_{\alpha\beta}\big)
\nonumber\\&
	\times\Bigg\{\theta\big(\nu^-_nE-(-1)^{n}eV-E_\alpha+E_\beta-\Delta\big)f\big(\nu^+_nE\big)f\big(\nu^-_nE-(-1)^{n}eV-E_\alpha+E_\beta\big)\frac{\nu^-_nE-(-1)^{n}eV-E_\alpha+E_\beta}		{\sqrt{\big(\nu^-_nE-(-1)^{n}eV-E_\alpha+E_\beta\big)^2-|\Delta|^2}}
\nonumber\\&
	+\theta\big(\nu^-_nE+(-1)^{n}eV+E_\alpha-E_\beta-\Delta\big)f\big(\nu^-_nE\big)f\big(\nu^+_nE-(-1)^{n}eV-E_\alpha+E_\beta\big)\frac{\nu^-_nE+(-1)^{n}eV+E_\alpha-E_\beta}{\sqrt{\big(\nu^-_nE+(-1)^{n}eV+E_\alpha-E_\beta\big)^2-|\Delta|^2}}
\Bigg\}
\label{eq:SCSCspinexcurrent}
\end{align}
\end{widetext}
for spin-exchange tunneling, where $\theta(x)$ is the Heaviside step function, $\Delta_\bfp=\Delta_\bfk=\Delta$, and the superconducting phase difference $\phi=0$. The sign-alternating coefficients $\nu^+_n=(-1)^{(n^2+n+2)/2}$ and $\nu^-_n=(-1)^{n(n+1)/2}$ change sign for every other term starting with plus and minus, respectively. 

\begin{figure}[b]
\begin{center}
\includegraphics[width=\columnwidth]{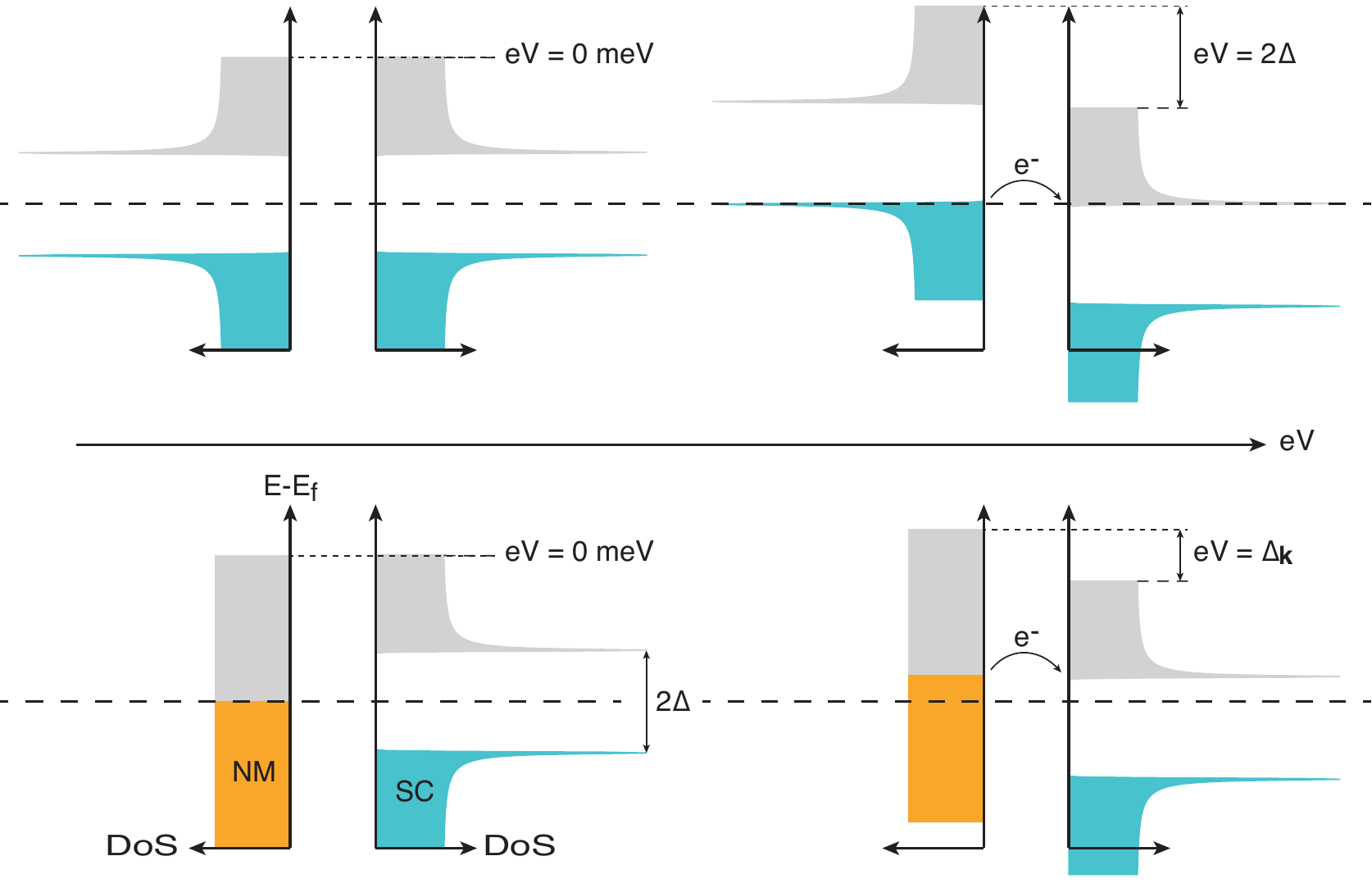}
\caption{(Color online) The bottom panel illustrates that the tunneling current onset happens at a bias voltage of $V=\Delta/e$ between a NM (orange) and a SC (turquoise). The DOS peak at onset carries over to the conduction spectrum. The top panel illustrates the same thing for a junction of SC:s where a bias voltage of twice the gap, $\Delta$, is needed for conduction.}
\label{fig1}
\end{center}
\end{figure} 

Despite the apparent added complexity, (\ref{eq:SCSCspinexcurrent}) behaves in much the same way as (\ref{eq:spinexcurrent}) with some qualitative differences. The onset of current by the applied bias voltage no longer happens when $eV\approx\Delta$, but instead occurs when $eV\approx2\Delta$ since the step functions include an additional pair potential, $\Delta$, to the lower integration limit. The additional fractions and step functions in the mathematical expressions for the currents also cause much sharper peaks in the $dI/dV$ spectra in comparison with the NM tip setup.

\section{Conduction spectra and analysis}
\label{sec-analysis}

\begin{figure*}
\includegraphics[width=\textwidth]{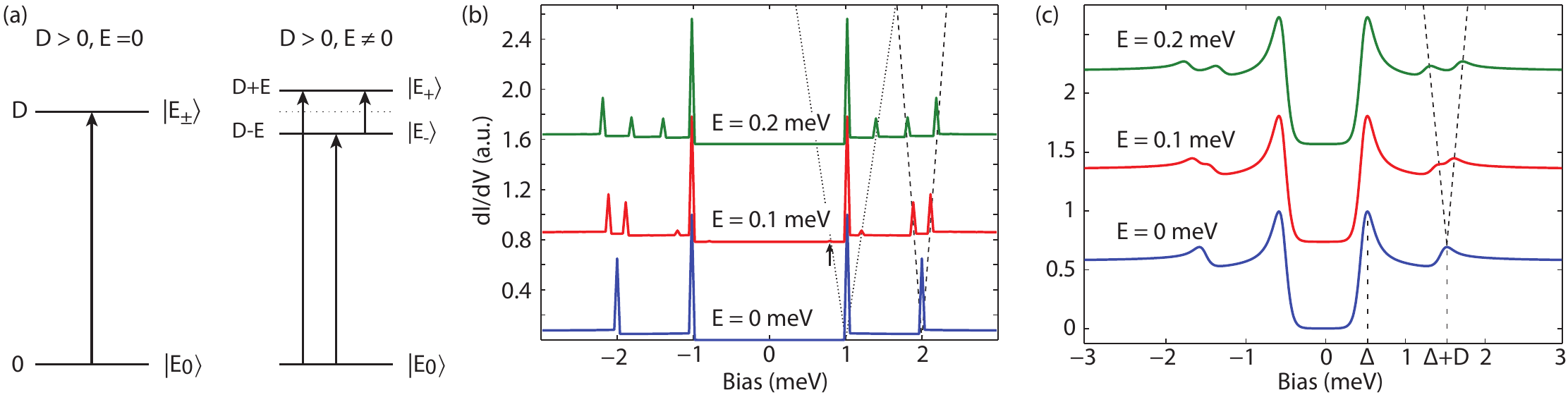}
\caption{(Color online) (a) Schematic picture of the possible spin transitions for $S=1$ with 0 or finite transverse anisotropy $E$. (b) Calculated conductance spectra of a SC-SC junction for local spin moment $S=1$ at $T=1.2$ K with parameters $D=1$ meV, $\Delta_{tip/sub}=0.5$ meV, $T_1=0.3T_0$, and varying $E$. (c) Same as in (b) for a NM-SC junction at $T=0.5$K.}
\label{fig2}
\end{figure*}

In contrast to a STM setup with normal-metal leads, the use of superconductors brings two main characteristic differences to the $dI/dV$ spectra that we have touched upon. First, tunneling electron-induced spin excitations that are energetically within the superconducting gap of the system never occur until the bias voltage has passed the gap. Energy exchange between tunneling electrons and the local spin moment is, in other words, shifted to $|V|=(\Delta+E_{\beta}-E_\alpha)/e$, for a NM-to-SC junction and $|V|=(2\Delta+E_{\beta}-E_\alpha)/e$ for a SC-to-SC junction, rather than $|V|=(E_{\beta}-E_\alpha)/e$, for a NM to NM junction. Though mathematically intrinsic, the physics picture to bear in mind is that tunneling from a SC at low temperatures only happens once enough energy is available to break up one Cooper pair. When tunneling to a SC, a single electron cannot occupy a SC low-lying state but must find a quasiparticle state higher in energy. The minimum energy cost for either event is $\Delta\hspace{1mm}\mbox{eV}$ when one lead is a SC and $2\Delta\hspace{1mm}\mbox{eV}$ when two SC leads are used. Second, while inelastic scattering signatures in a NM-to-NM junction, of leads with flat density of states (DOS), appear as steps of increased conduction at the onset energies in the $dI/dV$ spectra,\cite{balashov2009, hirjibehedin2006, hirjibehedin2007, khajetoorians2010} SCs produce peak structures  followed by the usual stepped increase. These peak structures are left in the $dI/dV$ curve as a trace by the underlying SC DOS, which exhibit pronounced coherence peaks at the end of the gap on both the occupied and unoccupied side.\cite{balatsky2003} Just as the bias voltage provides enough energy for an additional conduction channel to open, either the occupied or unoccupied states are inevitably at peak density. The conduction is momentarily high and falls off once the bias has passed the peak. See Fig. \ref{fig1} for an illustrative description of these tunneling properties. 

\subsection{Spin 1 magnetic molecule}
Three quantum states $|m_z=-1,0,1\rangle$ exist for a local magnetic moment of $S=1$ and a finite axial anisotropy $D$ generates two energy eigenvalues to the spin Hamiltonian $\Hamil_S$ if the transverse field $E$ and the external magnetic field $B$ are absent. A positive anisotropy $D>0$ meV will cause the eigenstates $|m_z=\pm1\rangle$ to lie $E_{\pm}=D$ meV above the state $|m_z=0\rangle$, thus favoring a low-spin ground state. A negative anisotropy $D<0$ meV will, instead, favor the high-spin state since $E_\pm<E_0$.

The left panel in Fig. \ref{fig2}(a) schematically shows the possible spin transitions in the case $D>0$. Hence, feeding the energy corresponding to $|D|$, in addition to the energy needed to overcome the superconducting gap(s), into the system allows the local spin moment to undergo transitions between its ground and excited states, which is clearly illustrated in the bottom traces of Figs. \ref{fig2}(b) and \ref{fig2}(c). Here, additional conductance channels emerge at $eV=\pm2|\Delta|+D$ in the SC-SC junction and $eV=\pm|\Delta|+D$ in the NM-SC junction, respectively, caused by the inelastic scattering. In this case, we have used the values $\Delta=0.5$ meV and $D=1$ meV. Qualitatively, the conductance spectrum in Fig. \ref{fig2}(b) agrees well with the experimentally obtained conductance in Ref. \onlinecite{franke2011}.

The $dI/dV$ curves of the SC-SC and the NM-SC junctions in Figs. \ref{fig2}(b) and \ref{fig2}(c) are in stark contrast to each other in terms of peak width. The SC-SC peaks are very sharp even though the calculations were done at a temperature of $1.2$ K as opposed to $0.5$ K for the NM-SC case. The difference is to be expected to some extent since two DOS coherence peaks match up at the onset of any new conduction channel to give a very conductive SC-SC junction for a narrow voltage span. In contrast, the conduction for a NM-SC junction, where one lead has a flat DOS, differs less at onset voltage in comparison to higher voltages. While this reasoning will explain a noticeable difference, the huge discrepancy found in our calculations indicates a failure of theory to handle peak widths in the SC-SC situation.


For $E\neq0$, the eigensystem of the local spin is modified to $E_\pm=D\pm E$, $|E_{\pm1}\rangle\equiv[|m_z=-1\rangle\pm|m_z=1\rangle]/\sqrt{2}$, which breaks the degeneracy and separates $|E_{+1}\rangle$ from $|E_{-1}\rangle$ by $2E$ in energy. The spin changing transitions, e.g., $\langle E_0|S_+|E_-\rangle$ and $\langle E_0|S_-|E_+\rangle$, therefore occur at different energies, as illustrated in the right panel of Fig. \ref{fig2}(a), and we expect conductance signatures at the voltage biases $|eV|=\Delta_{\rm sub}+D\pm E$ for the NM-SC setup and at $|eV|=\Delta_{\rm tip}+\Delta_{\rm sub}+D\pm E$ in the SC--SC case, which is readily seen in Figs. \ref{fig2}(b) and \ref{fig2}(c), respectively. The dashed and dotted lines trace the actual progression of eigenvalue differences with respect to increasing values of the transverse anisotropy. In addition, because the Fock states $|m_z=\pm1\rangle$  are coupled, the tunneling current also facilitates spin-preserving transitions between the states $|E_{+1}\rangle$ and $|E_{-1}\rangle$. Inelastic signatures between these higher-energy spin states are expected to appear on both sides of the main coherence peaks. At $|eV|=\Delta_{\rm tip}+\Delta_{\rm sub}+E$, the dotted line leaning towards the right in Fig. \ref{fig2}(b) traces peaks from excitations $|E_-\rangle\rightarrow|E_+\rangle$. The dotted line leaning towards the left traces the barely visible in-gap peaks, indicated by an arrow in the middle curve, at $|eV|=\Delta_{\rm tip}+\Delta_{\rm sub}-E$ from deexcitations $|E_+\rangle\rightarrow|E_-\rangle$ that assist electrons in tunneling. The higher-energy states of the local spin reveal themselves in this manner since they are thermally populated enough at $k_BT\propto0.1{\rm meV}$ to support transitions. The NM-SC $dI/dV$ curves of Fig. \ref{fig2}(c) are calculated at a lower temperature that populates the higher-spin states less, which in turn prevents a clear signature from $|E_{+1}\rangle\leftrightarrows|E_{-1}\rangle$ transitions. 

The apparent difference in amplitude between the transitions $|E_0\rangle\rightarrow|E_{\pm1}\rangle$ and $|E_{\pm1}\rangle\rightarrow|E_{\mp1}\rangle$, which is legible from Fig. \ref{fig2}(b), can be understood in terms of the population factors $P_{\alpha\beta}$. For $D>0$ and small $E\neq0$, the populations $P_\pm$ of the states $|E_{\pm1}\rangle$ are both close to 0, such that, e.g., $(1-P_+)P_-$ becomes small. The population $P_0$ for the state $|E_0\rangle$ is, on the other hand, close to 1 which leads to relatively large products $(1-P_\pm)P_0$. Note also that as $P_-$ gets larger for greater values of $E$, while $P_+$ gets smaller, the $(1-P_+)P_-$ peak gets bigger. At the same time, $(1-P_-)P_0$ becomes smaller while $(1-P_+)P_0$ gets slightly bigger, even though the low initial value of $P_+$ prevents any considerable changes.

\subsection{Spin 5/2 magnetic molecule}
Next, we turn our attention to the spin $S=5/2$ system in order to connect to recent experimental observations.\cite{Heinrich2013} For $E=0$, the eigensystem consists of the doubly degenerate states $|m_z=\pm m/2\rangle$, $m=1,3,5$, at energies $E_{\pm m/2}=Dm^2/4$, and, with a positive (negative) uniaxial anisotropy, $D>0$ ($D<0$), the system acquires a minimal (maximal) spin state $|\pm1/2\rangle$ ($|\pm5/2\rangle$).

\begin{figure}[t]
\begin{center}
\includegraphics[width=0.99\columnwidth]{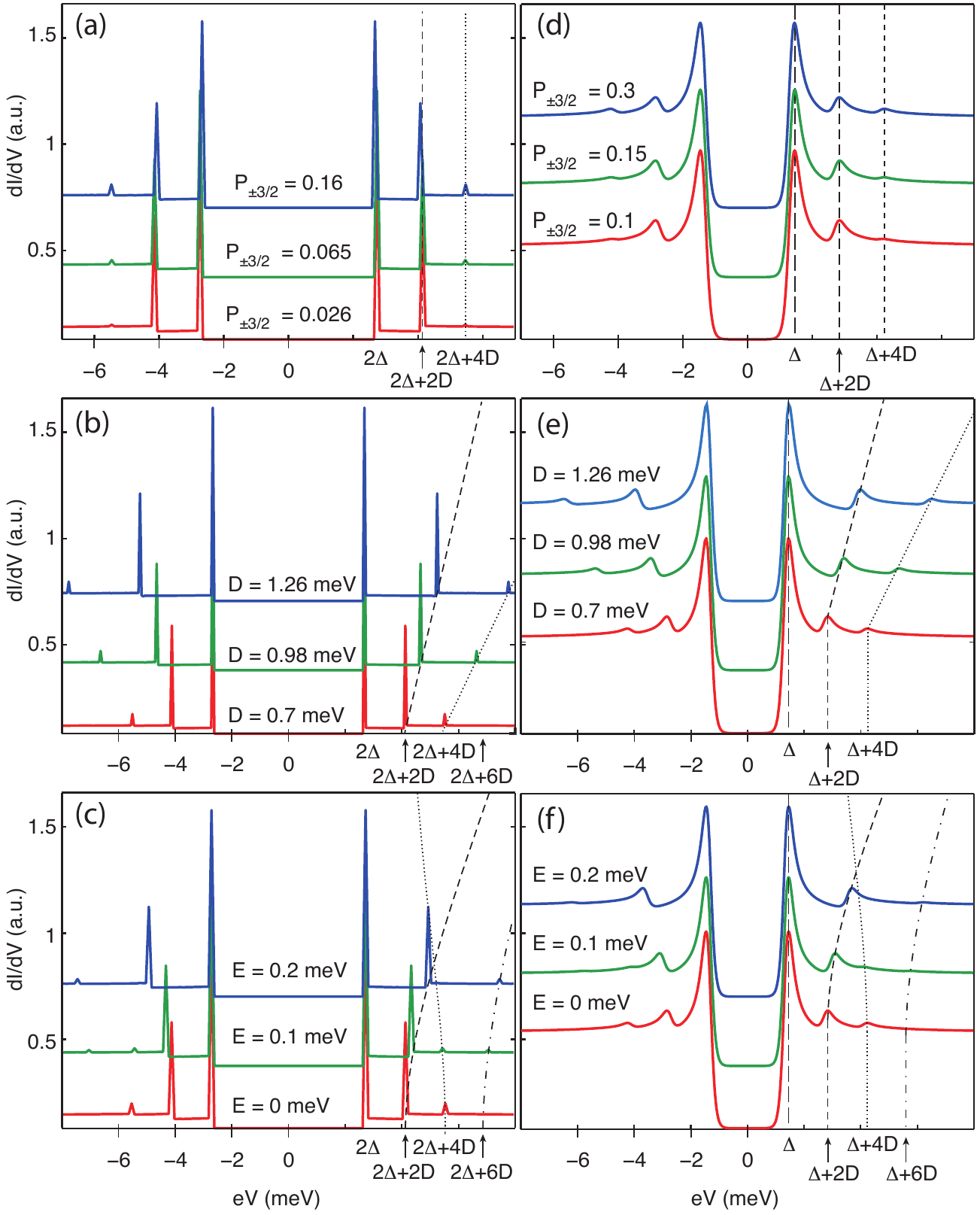}
\caption{(Color online) (a)--(c) Calculated SC-SC conductances for a spin $S=5/2$ system under (a) varying  population of the states $|m_z=\pm3/2\rangle$ for $D=0.7$ meV, $E=0$, (b) varying uniaxial anisotropy $D$, for $E=0$, and (c) varying transverse anisotropy $E$, where $D=0.7$ meV and $P_{\pm3/2}=0.16$. Other parameters are $\Delta_{\rm tip/sub}=1.35$ meV, $T=1.2$ K,\cite{Heinrich2013} and $T_1=0.3T_0$. (d)--(f) Corresponding conductance spectra for a NM-SC junction for parameter values $T=1$ K and $T_1=0.1T_0$, while other parameters are as in (a)--(c).}
\label{fig-S2}
\end{center}
\end{figure}

In Fig. \ref{fig-S2}(a), we plot the calculated SC-SC junction conductance for varied populations of the states $|m_z=\pm3/2\rangle$ in the absence of transverse anisotropy, $E=0$. We infer that our model calculations reproduce the experimental observations with excellent agreement. Here, we assume that the pairing potentials of the tip and substrate are equal, $\Delta_{\rm tip/sub}=\Delta\sim1.35$ meV, neglect possible superconducting phase differences, and use a positive uniaxial anisotropy $D=0.7$ meV. Analogously to the previous case, the conductances display strong coherence peaks at $eV=\pm2\Delta$, which are perfectly replicated at the voltage biases $|eV|=2\Delta+2D$ for the inelastic spin transition $|m_z=\pm1/2\rangle\rightarrow|m_z=\pm3/2\rangle$.

We, furthermore, notice the conductance peak emerging at voltage biases $|eV|=2\Delta+4D$ for an increased population of the first excited states $|m_z=\pm3/2\rangle$. The conductance peak is a signature of the inelastic transition $|m_z=\pm3/2\rangle\rightarrow|m_z=\pm5/2\rangle$ and its characteristics can be quantified by using the expressions in Eqs. (\ref{eq:spincorrel1}) and (\ref{eq:spincorrel2}). As the matrix elements for raising and lowering between the states $|m_z=\pm3/2\rangle$ and $|m_z=\pm5/2\rangle$ are always finite in the present setup, the emergence of the conductance peak strongly depends on the population of these states. When the ground state is heavily populated, both $|m_z=\pm3/2\rangle$ and $|m_z=\pm5/2\rangle$ are largely unpopulated and the factors $P_{\pm\frac{3}{2}\pm\frac{5}{2}}$ are vanishing. This scenario remains valid for small charge currents through the system, as well. For increasing charge currents, however, population density is expected to accumulate in the states $|m_z=\pm3/2\rangle$ as they are excited with a faster rate than their corresponding decoherence times. Accordingly, upon populating these states, the factors $P_{\pm\frac{3}{2}\pm\frac{5}{2}}$ become finite which leads to the transitions $|m_z=\pm3/2\rangle\rightarrow|m_z=\pm5/2\rangle$ contributing additional channels for conduction. In this fashion, we reproduce the effect of pumping which is obtained experimentally by decreasing the distance between the scanning tip and the sample.

Figure \ref{fig-S2}(d) illustrates the corresponding conductance spectra for a spin-5/2 magnetic molecule trapped within the gap of a NM-SC junction. Once again, no qualitative differences are obvious from the SC-SC case, except for wider peaks and earlier onset, at bias voltages $|eV|=\Delta$ for the main conductance peak and at $|eV|=\Delta+2D$ for the $|m_z=\pm1/2\rangle\rightarrow|m_z=\pm3/2\rangle$ transition. With higher population numbers of the state $|m_z=\pm3/2\rangle$, which are motivated if the local spin mainly dispenses excitation energy and angular momentum to the SC substrate through the relatively slow spin-phonon coupling to allow for pumping, signatures from the inelastic $|m_z=\pm3/2\rangle\rightarrow|m_z=\pm5/2\rangle$ transition are revealed.\cite{leuenberger1999} 

The plots in Fig. \ref{fig-S2}(b) and \ref{fig-S2}(e) show the evolution of the IETS spectra as a function of the effective uniaxial anisotropy $D$, which may be thought of as the sum of the intrinsic molecular anisotropy and the Cooper-pair-induced anisotropy; cf. the model in Eq. (\ref{eq-effHs}). The shift to higher energies of the inelastic peaks is expected from the previous discussion in Sec. \ref{sec-modifications}. In the experiment, the STM tip was brought closer to the sample which is expected to generate an exponential growth of the tunneling current since the coupling between tip and sample varies exponentially with distance. Here, as we do not attempt to model the whole experimental setup, but rather investigate the effects of changes in the anisotropy, we have not included this exponential variation of the anisotropy. In the case of a NM-SC junction, we may think of the tip as superconducting while the substrate is normal metallic for a feasible physical setup where an increase of the effective $D$ follows when the STM tip is brought closer to the sample.

For a finite transverse anisotropy, $E\neq0$, a peak can be seen to rise along the dash-dotted line in the SC-SC panel of Fig. \ref{fig-S2}(c) as the value of $E$ gets bigger. To explain the appearance of this peak, we look at how the spin states modify simultaneously to form linear combinations of the kind $|E_{\pm m}\rangle=\sum_{n=1,3,5}\alpha_{\pm n/2}^{(m)}|m_z=\pm n/2\rangle$. The six spin states are still doubly degenerate on three energy levels, but there is now a finite probability that a transition from the lowest state, e.g., $|E_{1}\rangle$ weighted on $|m_z=1/2\rangle$, to the highest, e.g., $|E_3\rangle$ weighted on $|m_z=5/2\rangle$, occurs despite seemingly violating conservation of angular momentum. Consequently, increased values of $E$ distribute density among the Fock states to allow for transitions with $\Delta m_z=\pm1$ between any of the available states. A schematic picture of the added transition possibilities for nonzero transverse anisotropy $E$ is given in Fig. \ref{fig-S3}. With different values of $E$, the spin state energy levels also shift relative to each other, which is reflected in the peak positions of Fig. \ref{fig-S2}(b). For example, at just over $E\approx0.2\hspace{1mm}{\rm meV}$, e.g., the $|E_1\rangle\rightarrow|E_2\rangle$ and $|E_2\rangle\rightarrow|E_3\rangle$ transitions clearly cross in energy.  

The characteristics of the SC-SC conductance spectra translate, once again, to the NM-SC case for finite values of $E$ since both systems share the local spin structure; see Fig. \ref{fig-S2}(f). Spectral details of the internal workings are, however, easily lost in the thermal broadening of the transition peaks.

\subsection{Influence of the temperature}

\begin{figure}[t]
\begin{center}
\includegraphics[width=0.99\columnwidth]{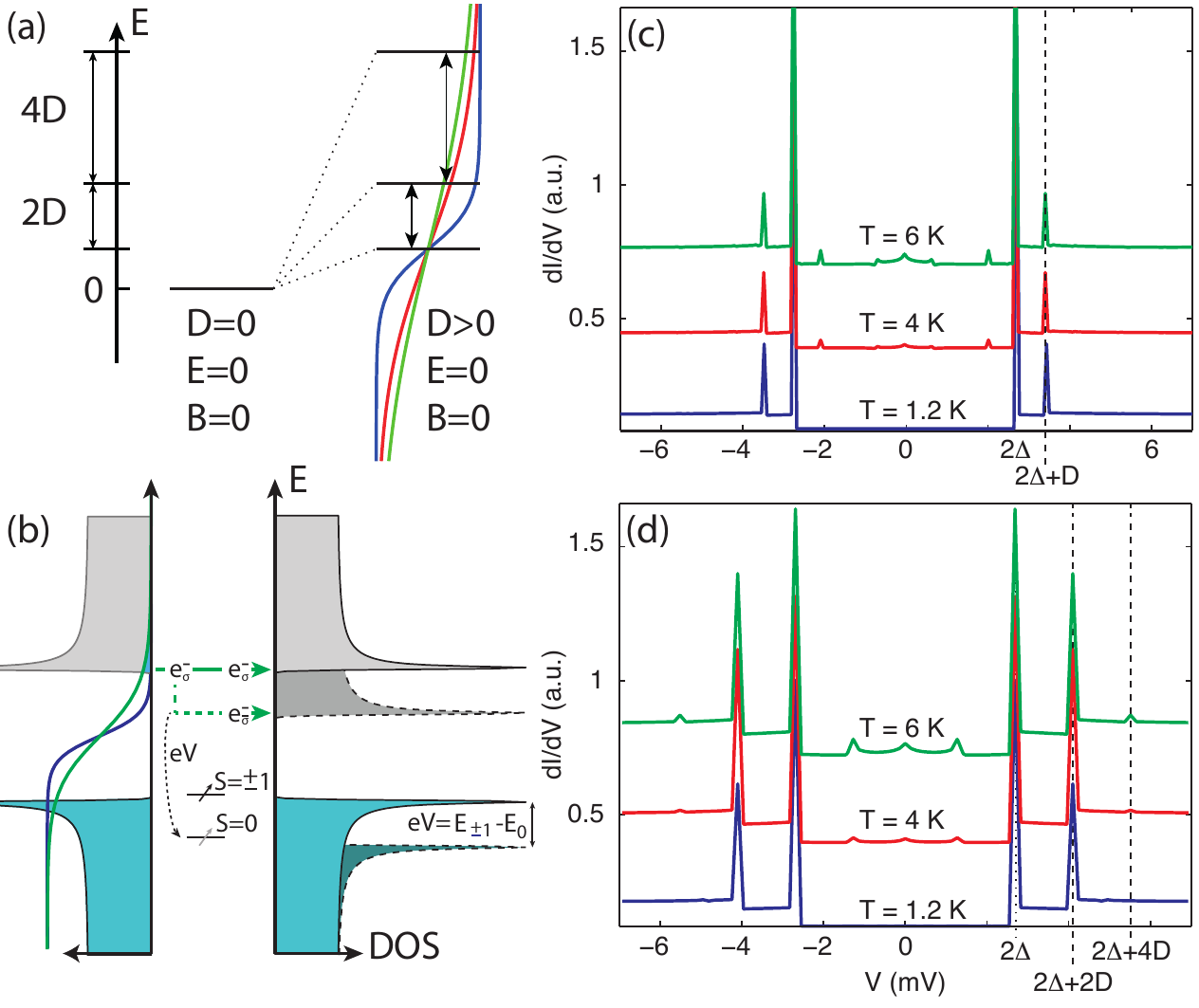}
\caption{(Color online) (a) Schematic spectrum of a spin $S=5/2$ SC-SC system in the atomic limit subject to different conditions. The blue, red, and green traces in the right panel represent the Fermi function at temperatures $T=1.2/4/6$ K. (b) Schematic DOS illustration depicting temperature effects for a $S=1$ system. (c), (d) Conductance spectra for the SC-SC setup at temperatures corresponding to those given in (a) and (b) for a $S=1$ and $S=5/2$ system, respectively. Other parameters are as in Fig. \ref{fig-S2}.}
\label{fig-S52T}
\end{center}
\end{figure}

We notice in Eq. (\ref{eq:SCSCspinexcurrent}) that there is an increased degree of detail in the tunneling current and, hence, the conductance spectra, at elevated temperatures, provided that we remain below the critical temperature. In Figs. \ref{fig-S52T}(c) and \ref{fig-S52T}(d), these details are reflected in the conductance spectra as additional peaks that rise with increased population of states higher in energy. For the in-gap peaks, there are two mechanisms responsible, which are of similar origin.

To begin with, consider the $S=5/2$ system, whose spectrum for finite $D>0$ is provided schematically in Fig. \ref{fig-S52T}(a). For low temperatures and small currents, the excited states are expected to be more or less unoccupied, while the ground state, covered by a finite portion of the Fermi function (blue), is occupied. Under these conditions, we retain the previous conductance spectrum, which is reproduced in the bottom trace of Fig. \ref{fig-S52T}(d) for convenience.

\begin{figure*}[t]
\begin{center}
\includegraphics[width=0.99\textwidth]{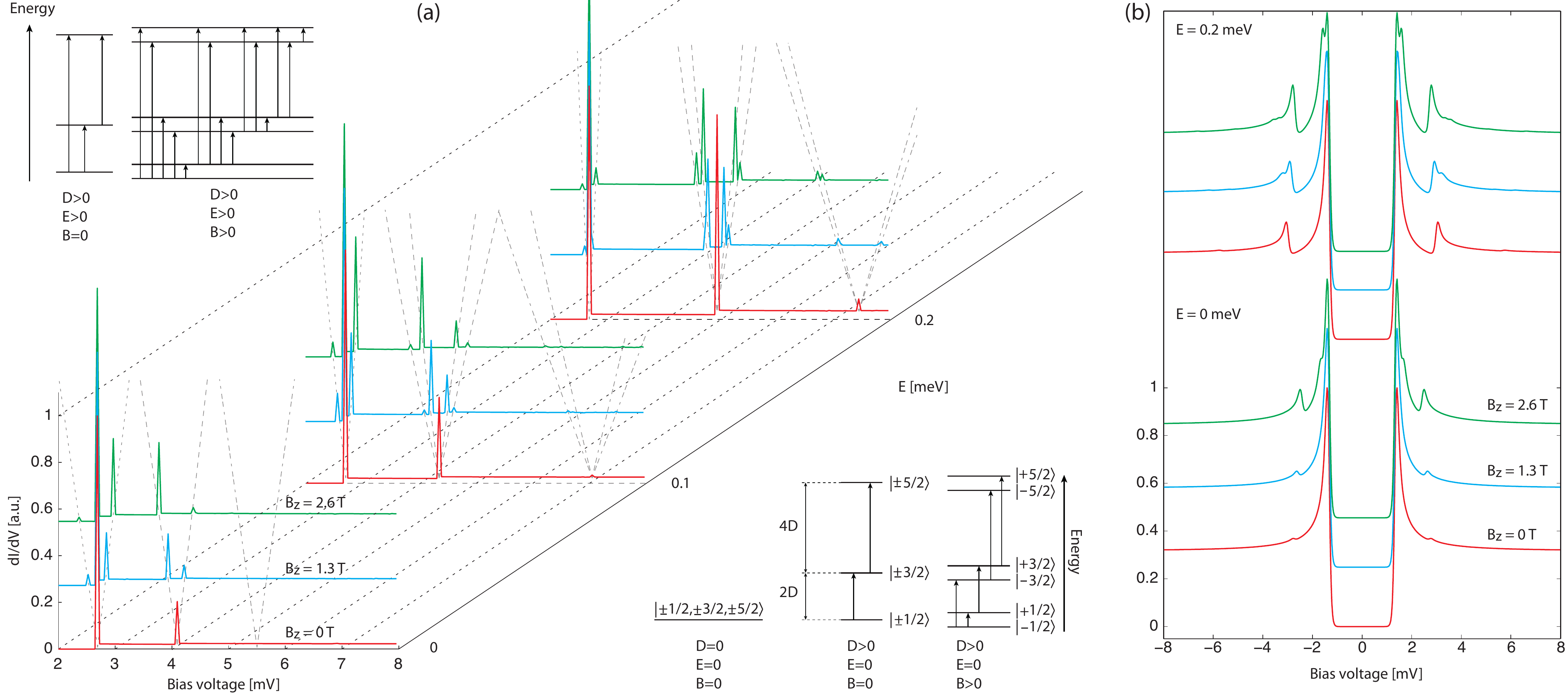}
\caption{(Color online) (a) Spectrum of a spin $S=5/2$ SC--SC system in the atomic limit subject to different conditions, parametrized by the anisotropies $D, E$ and external magnetic field ${\bf B}=B\hat{\bf z}$. (b) Corresponding conductance spectra for a NM-SC setup. Unspecified parameters are as in Fig. \ref{fig-S2}.}
\label{fig-S3}
\end{center}
\end{figure*}

At elevated temperatures, the thermal excitation energy is sufficient for the excited states to be partially occupied; see Fig. \ref{fig-S52T}(a), red and green Fermi functions. The local spin can then undergo transitions not only from lower to higher excitations, but also from higher to lower. In the former case, the spin moment has to absorb energy from the tunneling current, hence, the voltage bias has to be sufficiently large to assist such a transition, e.g., $|eV|\leq2|\Delta|+E_{3/2}-E_{1/2}$, where $E_{3/2}-E_{1/2}>0$. In the latter case, however, the spin moment is already thermally excited and can undergo deexcitation processes at energies, e.g., $|eV|\leq2|\Delta|+E_{1/2}-E_{3/2}$, where $E_{1/2}-E_{3/2}<0$. Then, the spin moment emits the energy quanta $E_{3/2}-E_{1/2}$ into the tunneling current, a process that opens a new channel for conduction which is expected to be seen within the gap of the conductance spectrum. This is indeed the case, evident in the conductance traces calculated for $T=4$ and $T=6$ K in Fig. \ref{fig-S52T}(d). The peak near equilibrium corresponds to the transition $\ket{\pm3/2}\bra{\pm5/2}$ since $4D=2.8$ meV $\sim2|\Delta|=2.7$ meV. Given that the thermal excitation energy $k_BT$ is greater than the difference between the two energies, $\approx0.3$ and $0.5$ for $T=4$ and 6 K, respectively, there is room for these deexcitation processes near equilibrium. The second in-gap feature corresponds to the emission resonance for the transition $\ket{\pm1/2}\bra{\pm3/2}$, emerging equidistantly from the superconducting coherence peak as its corresponding absorption resonance.

While spin deexcitation assisted tunneling certainly accounts for some of the in-gap features, we continue our discussion by looking at the spectra for S=1 given in Fig. \ref{fig-S52T}(c). Only one excitation peak exists alongside its deexcitation signature close to the main coherence peak. Yet, there is still an emerging three-peak structure forming with higher temperatures in the center of the superconducting gap. A second mechanism is clearly at play and it can be explained by looking at the schematically drawn DOS in Fig. \ref{fig-S52T}(b). For low temperatures, the Fermi function (blue) occupies the subgap states only, but for higher temperatures, the Fermi function (green) stretches all the way to the overgap states that become slightly filled. The conductance is consequently nonzero for low-bias voltages through direct spin-preserving tunneling. Once the voltage bias passes the $E_{\pm1}-E_0$ difference, spin-flip tunneling, which excites the local spin, may also occur, adding a peak at $eV=D$ meV. In-gap resonances were recorded for a $S=1$ Mn-phthalocyanine at $T=4.5$ K, \cite{franke2011} and we believe that the in-gap resonances considered here can, at least partly, explain these observations.

\subsection{Spin 5/2 magnetic molecule under external magnetic field}
In order to explore additional aspects of the conduction spectra for the $S=5/2$ system, an external magnetic field is introduced to break up the twofold degeneracies that the anisotropy fields $D$ and $E$ are unable to. Figure \ref{fig-S3}(a) pictures the expected behavior for three different magnetic field intensities in the $z$- direction. For $E=0$, two smaller peaks emerge around the main coherence signature, at $eV=2\Delta$, equidistant on both sides as a result of inelastic emission and absorption between the Zeeman split ground states, $|m_z=-1/2\rangle\leftrightarrows|m_z=1/2\rangle$.  Forking off as the magnetic field increases at about $V=4\hspace{1mm}{\rm mV}$  are two peaks that signal transitions between $\langle m_z=3/2|S_+|m_z=1/2\rangle$ and $\langle m_z=-3/2|S_-|m_z=-1/2\rangle$. These transitions differ in energy because the pair of ground states are Zeeman split by a different amount than the first excitation pair of states. Note that the transitions occur between uncoupled basis states since $E=0$, which limits the number of possible excitation paths to 5, as schematically illustrated in Fig. \ref{fig-S3}. Indications of transitions between the first and second pair of excitation states are absent unless the effects of pumping are replicated as done in the previous example with no external magnetic fields.   

For $E>0$, the basis states once more couple to form eigenstates to the spin Hamiltonian. This is reflected in the $dI/dV$ plots of Fig. \ref{fig-S3}(a) for increased magnetic fields in the $z$-direction as a branch off of the transition signature, at $V\sim4\hspace{1mm}{\rm mV}$, into four peaks rather than the previous two for $B_z=0$. Any transition $|E_{\pm1}\rangle\rightarrow|E_{\pm2}\rangle$ is hence sufficiently probable to yield a visible peak in the conductance spectra. For $E\sim0.2\hspace{1mm}{\rm meV}$, we even begin to see four distinct peaks split off, in step with the magnetic field, that originate from excitations between the ground and the second excited states  at $V\sim7\hspace{1mm}{\rm mV}$. In theory, transitions are now allowed between all  spin states of the magnetic molecule, once $D>0$, $E>0$, and $B_z>0$, even though thermal populations for all but the two lowest-energy states are so small that excitations from higher states are rare occurrences; see upper left corner of Fig. \ref{fig-S3} for a diagram of the possible excitations.  

In Fig. \ref{fig-S3}(b) we look at the system under equal circumstances regarding the external magnetic field for the NM-SC setup. Unfortunately, reasonable magnetic fields separate the peaks from different spin transitions less than the thermal width which somewhat obscures details. Features of the underlying peak structure can still be made out as additional humps form with stronger magnetic fields, but aside from resolving in energy, it is possible to draw conclusions based on amplitude. For $E=0{\rm meV}$, a single peak appears to form at just over $V\sim2\hspace{1mm}{\rm mV}$ with an amplitude that is strongly dependent on the magnetic field. What appears to be one signature is really two peaks that separate for stronger fields. The peak moving towards the right, originating from $|m_z=1/2\rangle\rightarrow|3/2\rangle$, quickly dies off as $|m_z=1/2\rangle$ becomes less populated at the low temperature, while the peak moving towards the left, from the excitation $|m_z=-1/2\rangle\rightarrow|-3/2\rangle$, gains amplitude as $|m_z=-1/2\rangle$ becomes more populated. In this way, an external magnetic field can assist to increase amplitude for some transitions. 

\begin{figure}[t]
\begin{center}
\includegraphics[width=0.99\columnwidth]{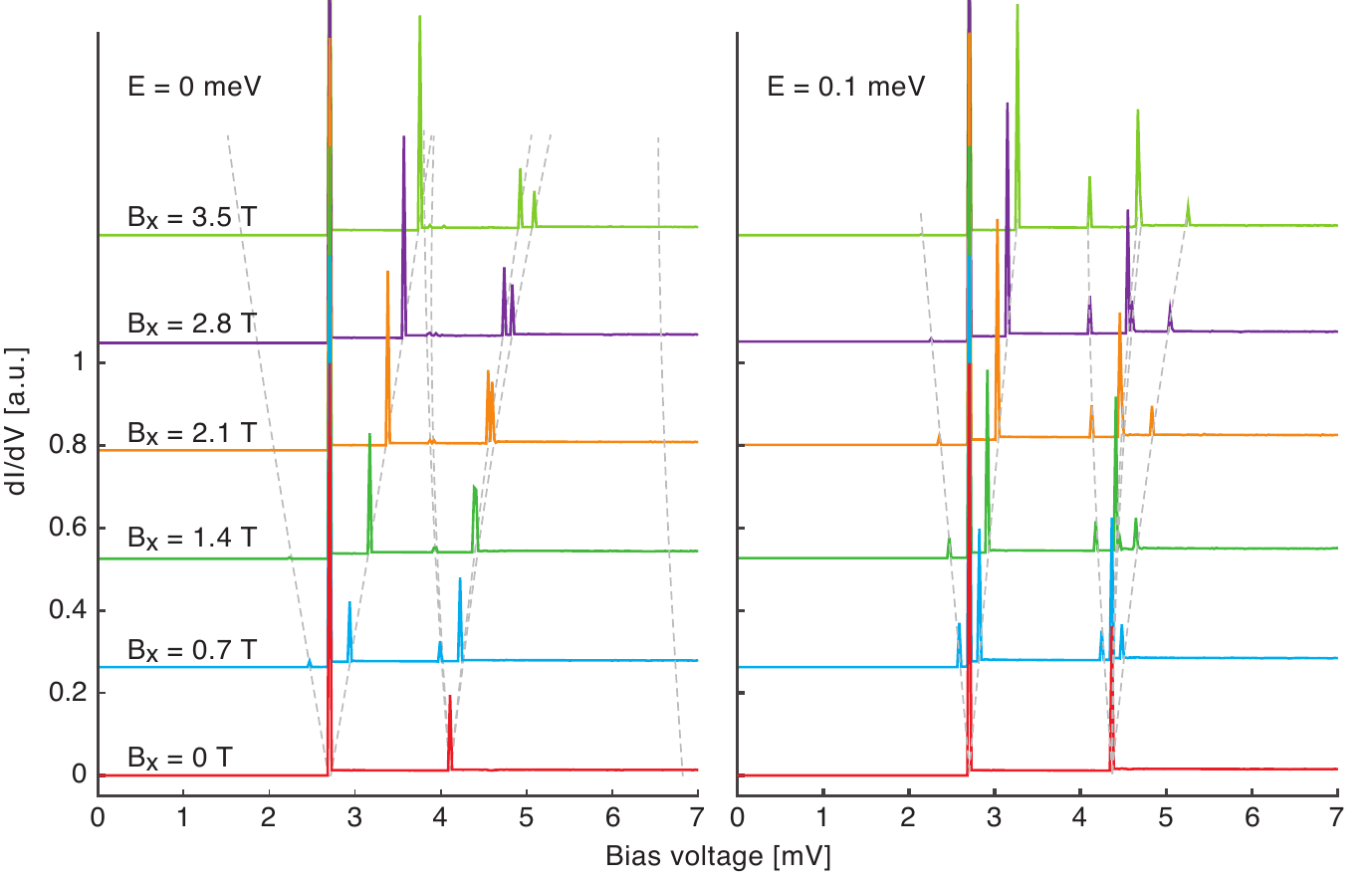}
\caption{(Color online) dI/dV spectra of a $S=5/2$ system under external magnetic fields in the $x$- direction with increasing field density. Only the transverse anisotropy $E$ differs between the two diagrams.}
\label{fig-S4}
\end{center}
\end{figure}

A magnetic field in the $x$- and $y$- directions similarly splits up the degenerate energy levels of the local spin. Around the main coherence peak, signatures from both absorption and emission can be seen when the spin leaps in energy between the separated ground states. Starting at approximately $V\sim4{\rm mV}$, we see in Fig. \ref{fig-S4} how the excitations $|E_{1(-1)}\rangle\rightarrow|E_{2(-2)}\rangle$, which share energy, produce a peak that divides into four when all transitions $|E_{1,-1}\rangle\rightarrow|E_{2,-2}\rangle$, at different energies, are allowed with the magnetic field, even though $E=0{\rm meV}$. In comparison with the $B_z\neq0$ setup, these peaks break apart along a bent path rather than following a straight line. The most frequent transitions are also those with higher energy as opposed to those with lower energy. When the transverse anisotropy $E$ is turned on, the conductance spectra look quite different as two peaks do not seem to separate, while the other two go off in opposing directions to effectively form a structure of three peaks.  

Providing an external magnetic field adds a complication to the measurements, since the superconductivity in both the substrate and tip becomes quenched under too strong fields. This problem can, however, be overcome by changing to a tip/substrate material that is less sensitive to magnetic fields, e.g., NbTi, Nb$_3$(Sn,Ge,Al), and MgB$_2$,\cite{larbalestier2001,buzea2001,gurevich2011} which are known to maintain their superconducting phase for fields as strong as 10--30 T. Our predictions made for fields up to a few T are therefore safely within the realm of feasibility.

\section{Summary and conclusions}
\label{sec-summary}
We argue that our simple model of a superconducting STM, holding a paramagnetic molecule within its gap, generates a differential conduction spectra that matches up very well to experimental data, taken of, e.g., Fe-OEP-Cl and Mn-phthalocyanine. The model notably captures peak signatures in the tunneling conductance from interactions with the local spin that reference to the sum of the tip and substrate pairing potentials rather than zero-bias voltage. We are also able to mimic the effects of electron pumping by introducing a uniform potential shift such that the excited spin states thermally populate to reveal peak imprints of transitions among them. The success up to these points leads us to infer that the key mechanism behind the experimental conductance features is exchange interaction between tunneling electrons and the local spin moment.

Our model does not include direct exchange between the local spin and the superconducting substrate, which will generate states within the superconducting gap, since we argue that the separating ligand cage weakens this interaction such that, e.g., Shiba states move close to the dominating coherence peak. Our model does, however, capture the effect of exchange between the local spin moment and the Cooper pair correlations that generates a finite contribution to the uniaxial anisotropy that acts on the local spin moment and increases with decreasing distance between the superconducting tip and the sample. This effect, therefore, offers an explanation for the increased anisotropy observed in experiments.

With the freedom to explore parameter space, we consider different magneto crystalline anisotropy values as well as the effects of an external magnetic field. For $S=5/2$, the axial anisotropy field directly determines level spacing between spin states, while the transverse anisotropy field, apart from slightly shifting the energy levels, couples the spin basis states to allow for transitions which are otherwise prohibited by conservation of angular momentum. An external magnetic field removes spin state degeneracies and provides a rich conductance spectrum.

We have also considered temperature effects up to the critical temperature and shown that both direct thermal excitations of the local spin, as well as thermal population of states above the superconducting gap, cause in-gap peaks in the conductance. Deexcitation of thermally populated higher-spin states assist electron tunneling at voltage biases lower than the superconducting gap, while thermally occupied states above the superconducting gap give a nonzero conductance contribution that peaks at zero voltage bias and when it matches up with the excitation energy of the local spin. These mechanisms may partially explain the observed in-gap resonance of Mn-phtalocyanine.
   
An extended experimental study of the system could benefit from the use of an external magnetic field. The main  argument for long spin excitation lifetimes is that deexcitations with an energy release between $0<\varepsilon<2\Delta$ fail to split up Cooper pairs and facilitate particle-hole creation. A magnetic field immediately produces a large peak that separates from the main coherence peak due to transitions between the no-longer degenerate ground states. The energy of this excitation varies with the strength of the magnetic field starting from $0{\rm meV}$ and upwards.

\acknowledgments
We thank A. Black-Schaffer and K. Bj\"{o}rnson for stimulating and fruitful discussions. This work was supported by the Swedish Research Council.

\bibliography{longspinartbib}
\end{document}